\documentclass[11pt]{article}
\usepackage{epsf}
\usepackage{graphicx}
\usepackage{latexsym}
\usepackage{amsfonts}
\usepackage{exscale}
\usepackage{epic}
\usepackage{eepic}
\newcommand{\eqnn}[1]{\begin{eqnarray*}#1\end{eqnarray*}}

\newcommand{\eqnl}[2]{\par\parbox{14cm}
{\begin{eqnarray*}#1\end{eqnarray*}}\hfill
\parbox{1cm}{\begin{eqnarray}\label{#2}\end{eqnarray}}}

\newcommand{\eqngrlb}[3]{\par\parbox{14cm}
{\begin{eqnarray}\fbox{$\displaystyle#1\\#2$}\end{eqnarray}}\hfill
\parbox{1cm}{\begin{eqnarray}\label{#3}\end{\eqnarray}}}

\newcommand{\eqngr}[2]{\begin{eqnarray*}#1\\#2\end{eqnarray*}}

\newcommand{\eqngrl}[3]{\par\parbox{14cm}
{\begin{eqnarray*}#1\\#2\end{eqnarray*}}\hfill
\parbox{1cm}{\begin{eqnarray}\label{#3}\end{eqnarray}}}

\newcommand{\eqngrrl}[4]{\par\parbox{14cm}
{\begin{eqnarray*}#1\\#2\\#3\end{eqnarray*}}\hfill
\parbox{1cm}{\begin{eqnarray}\label{#4}\end{eqnarray}}}

\newcommand{\refs}[1]{(\ref{#1})}

\newcommand{\be}{\begin{equation}}
\newcommand{\ee}{\end{equation}}
\newcommand{\ft}[2]{{\textstyle\frac{#1}{#2}}}
\def\ps{\!+\!}
\def\ms{\!-\!}
\def\lam{\lambda}
\def\al{\alpha}
\def\th{\theta}
\def\gam{\gamma}
\def\gamz{\gamma_0}
\def\gamo{\gamma_1}
\def\gamf{\gamma_5}
\def\an{\hat\lam}
\def\ns{n\hskip-2mm\slash}
\def\pa{\partial}
\def\ov{\over}
\def\id{1\!\mbox{l}}
\def\eth{e^{\theta}}

\def\erfc{{\rm erfc}}
\def\erf{{\rm erf}}
\def\mod{{\rm mod}}
\def\sign{{\rm sign}}
\def\Tr{{\rm Tr}}
\def\tr{{\rm tr}}
\def\mtxt#1{\quad\hbox{{#1}}\quad}
\def\an{a_n}
\def\Z{\mathbb{Z}}

\begin{document}

\begin{flushright}
La Plata Th 02-02\\FSUJ-TPI-02-02
\end{flushright}
\par
\vskip .5 truecm
\begin{center}
{\Large
Spectral asymmetry for bag boundary conditions}
\end{center}
\vskip 1truecm
\normalsize

\begin{center}
\textbf{
C.G. Beneventano, E.M. Santangelo}\footnote{e-mails:
gabriela, mariel@obelix.fisica.unlp.edu.ar}\\
\textit{Departamento de F\'{\i}sica, Universidad Nacional de La Plata\\
C.C.67, 1900 La Plata, Argentina}

\textbf{A. Wipf}\footnote{e-mail: wipf@tpi.uni-jena.de}\\
\textit{Theoretisch-Physikalisches Institut,
Friedrich-Schiller-Universit\"at Jena\\ D-07743 Jena, Germany}
\end{center} \par \vskip 1 truecm \begin{abstract} We give an expression,
in terms of boundary spectral functions, for the spectral asymmetry of the
Euclidean Dirac operator in two dimensions, when its domain is determined
by local boundary conditions, and the manifold is of product type. As an
application, we explicitly evaluate the asymmetry in the case of a
finite-length cylinder, and check that the outcome is consistent with our
general result. Finally, we study the asymmetry in a disk, which is a
non-product case, and propose an interpretation. \end{abstract}

\noindent PACS: 02.30.Tb, 02.30.Sa

\noindent MSC: 35P05, 35J55

\vskip .5 truecm


\section{Introduction }\label{section-0}

Spectral functions are of interest both in quantum field theory
and in mathematics (for a recent review, see \cite{habklaus}). In
particular, $\zeta$-functions of elliptic boundary problems are
known to provide an elegant regularization method \cite{zeta} for
the evaluation of objects as one-loop effective actions and
Casimir energies, as discussed, for instance, in the reviews
\cite{eliz}.

In the case of operators with a non positive-definite principal
symbol, another spectral function has been studied, known as
$\eta$-function \cite{gilkey}, which characterizes the spectral
asymmetry of the operator. This spectral function was originally
introduced in \cite{aps}, where an index theorem for manifolds
with boundary was derived. In fact, the $\eta$-function of the
Dirac operator, suitably restricted to the boundary, is
proportional to the difference between the anomaly and the index
of the Dirac operator, acting on functions satisfying nonlocal
Atiyah-Patodi-Singer (APS) boundary conditions. Some examples of
application were discussed in \cite{schroer,diskaps}.

Such nonlocal boundary conditions were introduced mainly for
mathematical reasons, although several applications of this type
of boundary value problems to physical systems have emerged,
ranging from one-loop quantum cosmology \cite{esp}, fermions
propagating in external magnetic fields \cite{bs} or so-called
$S-$branes, which are mapped into themselves under $T-$duality
\cite{vass}. So far, $\eta$-functions have found their most
interesting physical applications in the discussion of fermion
number fractionization \cite{nsem}: The fractional part of the
vacuum charge is proportional to $\eta(0)$. The $\eta-$function
also appears as a contribution to the phase of the fermionic
determinants and, thus, to effective actions \cite{cez}.
Furthermore, both the index and the $\eta$-invariant of the Dirac
operator are related to scattering data via a generalization of
the well-known Levinson theorem \cite{wirai}. A thorough
discussion of the index, $\zeta-$ and $\eta-$functions in terms of
boundary spectral functions for APS boundary problems can be found
in \cite{grubbseel,wojaps}.

Alternatively, one may consider the boundary value problem for the
Dirac operator acting on functions that satisfy local, bag-like,
boundary conditions. These conditions are closely related to those
appearing in the effective models of quark confinement known as
MIT bag models \cite{bag}, or their generalizations, the chiral
bag models \cite{chbag}. The physical motivation for studying
these local boundary conditions is thus clear.

In this paper, we will study the Euclidean Dirac operator in two
dimensions, acting on functions satisfying local bag-boundary
conditions \cite{balog,wipf}. Such boundary conditions are defined
through the projector in equation \refs{proj1} of the next
section. They contain a real parameter $\theta$, which is to be
interpreted as analytic continuation of the well known
$\theta$-parameter in gauge theories. Indeed, for $\theta\neq 0$,
the effective actions for the Dirac fermions contain a
$CP$-breaking term proportional to $\theta$ and proportional to
the instanton number \cite{wipf}; for example \eqngr{ \hbox{2
dimensions:}&&\Gamma_{\rm eff}\sim\theta\int d^2x F_{01}+\dots}
{\hbox{4 dimensions:}&&\Gamma_{\rm eff}\sim \theta\int
d^4x\epsilon_{\mu\nu\al\beta}F_{\mu\nu}F_{\al\beta}+\dots}

For $\th\neq0$ we will refer to the bag boundary conditions as
chiral while, in the particular case $\th=0$, we will call them
non-chiral or pure MIT conditions. In both cases, the Dirac
operator is self-adjoint. Moreover, in two dimensions, not only
the first order boundary value problem is elliptic, but also the
associated second order problem is so.

One of the main characteristics of bag boundary conditions is that
they lead to an asymmetry in the non-zero spectrum. Thus, in this
paper we will study the boundary contribution to the spectral
asymmetry for bag boundary conditions in two-dimensional Euclidean
space. The pure MIT case was studied, for any even dimension, in
\cite{woj3}. We will compare our results to those in this
reference whenever adequate.

Note that, as in any even dimension, there is no volume
contribution to the asymmetry (for a proof see, for instance,
\cite{gilkey}; qualitatively, this is due to the existence of
$\gamma_5$, which anticommutes with the Dirac operator). So, the
boundary contribution is also the total asymmetry. In section
\ref{section-2}, the asymmetry will be expressed in terms of
spectral functions of the boundary operator $A$. Throughout our
calculation in that section, we will assume the manifold to be of
product type near the boundary, and $A$ to be independent of the
normal variable.

As an example of a product manifold we will evaluate, in section
\ref{section-3}, the asymmetry in a finite cylinder with twisted
boundary conditions along the circle direction, imposing
APS-boundary conditions on one end of the cylinder and chiral bag
conditions on the other end. The result will be shown to be
consistent with our general prediction in section \ref{section-2}.

In section \ref{section-4}, we will compute the spectral asymmetry
in the case of a disk (two-dimensional bag), for chiral bag
boundary conditions. Note this is a non-product case; however, we
will suggest that the outcome of this calculation might be
understood from our general result in section \ref{section-2}.

Finally, section \ref{section-5} contains the generalization to
the case in which certain gauge potentials are present, as well as
some comments concerning the extension of our results to higher
dimensions.

\section{The heat kernel in terms of boundary eigenvalues }
\label{section-1}

In this section we rewrite the known heat kernel for the free
Euclidean Dirac operator on the semi-infinite cylinder subject to
bag-boundary conditions, such that the spectral resolution with
respect to the boundary operator becomes transparent. To this end,
it is convenient to choose a chiral representation for the
Euclidean $\gam$-matrices in $2$-dimensions, \eqnl{
\gamz=\sigma_1,\quad\gamo=\sigma_2\mtxt{and}
\gamf=-i\gamz\gamo=\sigma_3.}{gammatr} Then, the free Dirac
operator takes the form \eqnl{
P=i(\gamz\pa_0+\gamo\pa_1)=\left(\matrix{
  0&\pa_1 +A \cr
  -\pa_1 + A & 0 } \right),}{op}
where $A$ is the boundary operator \eqnn{ A=i\pa_0,} which will
play an important role in what follows. The euclidean
``time"-coordinate $0\leq x_0<\beta$ is tangential to the boundary
at $x_1=0$. The ``spatial" variable $x_1\geq 0$ is normal to the
boundary and grows toward the interior of the semi-infinite
cylinder. The projector defining the local bag boundary condition
\eqnn{ B\psi\big\vert_{x_1=0}=0} at the boundary $x_1=0$ reads
\eqnl{ B=\frac12(1-i\gamf e^{\gamf \th}\ns)=\frac12(1+i\gamf
e^{\gamf \th}\gamma_1) =\frac12\pmatrix{ 1&\eth  \cr
e^{-\th}&1},}{proj1} where $n^{\mu}$ is the outward oriented
normal, $n^{\mu}=(0,-1)$.

For convenience we introduce the variables
$\xi_\mu=x_\mu-y_\mu$ and $\eta=x_1+y_1$. Then, the heat kernel of
the associated second order operator reads, in terms of the
eigenvalues $\an $ of the boundary operator $A$,
\eqngrl{
K(t,x,y)&=&{1\ov \beta \sqrt{4\pi t}}\sum_n
e^{i\an \xi_0}e^{-\an ^2 t}
\Big\{e^{-\xi_1^2/4t}\id+ e^{-\eta^2/4t}M}
{&&\hskip 1.5cm -N\tanh\th e^{-\eta^2/4t}\Big[1-
{\sqrt{4\pi t}\ov \sinh 2\th}\,\an\,
\,e^{u_n(\eta,t)^2}\erfc\big[u_n(\eta,t)\big]\Big]\Big\}\,,}{proj2}
where we introduced the abbreviation
\eqnn{
u_n(\eta,t)={\eta\ov \sqrt{4t}}-\an\sqrt{t}\tanh\theta}
and the complementary error function,
\eqnn{
\erfc(x)={2\ov\sqrt{\pi}}\int_x^{\infty}dy\,e^{-y^2}.}
Moreover, $\id$ denotes the $2\!\times\!2$-identity matrix,
\eqnn{
M=\pmatrix{\eth \sinh{\th}&-\cosh{\th} \cr
   -\cosh{\th} & -e^{-\th}\sinh{\th}}\mtxt{and}
N=\pmatrix{\eth&-1\cr
   -1& e^{-\th}}\sinh\th .}
For finite temperature field theory, in which case the Dirac field
is antiperiodic in $x_0$ and
hence the eigenvalues of the boundary operator are
$\an =2\pi(n+1/2)/\beta$, the result \refs{proj2} coincides
with the Fourier transform of equation (101) in \cite{wipf2}.

\section{Boundary contribution to the spectral asymmetry
from bag boundary conditions}\label{section-2}

As already commented, since the euclidean space-time is even
dimensional, there is no bulk contribution to the asymmetry. To
obtain the boundary contribution, the eigenvalue problem for the
Dirac operator $P$ should be investigated on a collar neighborhood
of the boundary. Here, we consider instead the operator on the
semi-infinite cylinder extending to $x_1\rightarrow \infty$. As is
well-known \cite{woj3}, since we are treating a self-adjoint
problem, this yields the correct answer for an invertible boundary
operator $A$. We shall discuss the non invertible case toward the
end of this section. Hence, for the moment, we assume $\an\neq 0$
for all $n$.

Denoting the real eigenvalues of the Dirac operator by
$\lam$, the relevant spectral function is
\eqnl{
\eta(s,P)=\sum_{\lam}{\sign\lam\ov |\lam|^s}
=\zeta\Big(\frac{s+1}{2},P^2,P\Big)=
{1\ov \Gamma\big(\frac{s+1}{2}\big)}
\int_0^{\infty}dt\;t^{\frac{s-1}{2}}
\,\Tr \big(P\,e^{-tP^2}\big).}{eta1}

The Dirac trace can be computed with the help of \eqngr{
\tr(\gamma_{0,1}\id)=\tr(\gamma_1 M)\!&\!=\!&\!\tr(\gamma_1 N)=0
\mtxt{and}} {\tr(\gamma_0 M)=-2\cosh{\th}&,&\tr(\gamma_0
N)=-2\sinh{\th}\,.} From \refs{proj2} one obtains for the
Dirac-trace of the kernel needed in equation \refs{eta1} \eqngrl{
\tr\langle x\vert Pe^{-tP^2}\vert y\rangle&\!\!=\!\!& {\cosh\theta
e^{-\eta^2/4t}\ov i\beta\sqrt{\pi t}}\, \tr\Bigg({\pa\ov\pa
x_0}\sum_n e^{i\an \xi_0} e^{-\an^2 t}\Big\{1-\tanh^2\theta\cdot}
{&&\hskip3.5cm \cdot \Big[1-{\an \sqrt{4\pi t}\ov \sin 2\th}\,
e^{u_n(\eta,t)^2}\erfc\big[u_n(\eta,t)\big]\Big]\Big\}\Bigg)\,.}{t}
After performing the derivative with respect to $x_0$, setting
$x_\mu=y_\mu$ and integrating over the tangential variable, one is
left with the following integral over the normal variable
$x_1\equiv x$: \eqnl{ \Tr \big(Pe^{-tP^2}\big)=\sum_n\an e^{-\an
^2 t}\int\limits_0^{\infty}dx \Big\{{1\ov \sqrt{\pi t}}+
\an\tanh\th\, e^{u_n^2(2x,t)}\;\erfc\big[u_n(2x,t)\big]\Big\}\,
{e^{-x^2/t}\ov\cosh\th},}{t1} where we took into account that for
$x_\mu=y_\mu$ we have \eqnn{
u_n(\eta,t)=u_n(2x,t)={x\ov\sqrt{t}}-\an\sqrt{t}\tanh\theta,\qquad
x=x_1.} Now, we may use the simple identity \eqnn{ -{1\ov
2}{\pa\ov \pa x}
\Big[e^{-x^2/t+u_n^2(2x,t)}\erfc\big[u_n(2x,t)\big]\Big]
=e^{-x^2/t}\Big[{1\ov\sqrt{\pi t}}+\an \tanh\th\, e^{u_n^2(2x,t)}
\erfc\big[u_n(2x,t)\big]\Big]} to rewrite the relevant trace as
follows, \eqngrl{ \Tr \big(Pe^{-tP^2}\big)\!&=&\!-{1\ov 2\cosh\th}
\sum_n\an e^{-\an ^2 t/\cosh^2\th}\int_0^{\infty}dx {\pa\ov \pa
x}\left[e^{-2x\an \tanh\th} \erfc\left(u(2x,t)\right)\right]}
{&=&\!\frac12 \sum_n {\an\ov \cosh\th} e^{-\an ^2 t/\cosh^2\th}
\erfc\big[-\sqrt{t}\tanh{\th}\an \big]\,.}{trace} The asymmetry is
obtained by inserting \refs{trace} into \refs{eta1} and, hence, it
is given by \eqnn{ \eta(s,P) =\frac{1}{\Gamma(\ft{s+1}{2})}\sum_n
{\an \ov 2\cosh{\th}}\int_0^{\infty}dt\,t^{\ft{s-1}{2}} e^{-\an^2
t/\cosh^2\th} \left[1-\erf\big(-\sqrt{t} \tanh{\th}\an
\big)\right]\,,} where $\erf$ is the error function, \eqnn{
\erf(x)=1-\erfc(x)={2\ov\sqrt{\pi}}\int_0^{x}dy\,e^{-y^2}\,.}
Finally, changing variables to $\tau=\an ^2 t/\cosh^2{\th}$,
interchanging the order of the integrations and integrating over
$\tau$ one obtains the following rather explicit expression
\eqngrl{ \eta(s,P)&=&\frac12 \cosh^s\theta\sum_{n}
\big(\an^2\big)^{-s/2} \Big[\sign(\an)+I(s,\theta) \Big]} {&=&
\frac12\cosh^s\theta \Big[\eta(s,A)+
\zeta(\ft{s}{2},A^2)I(s,\theta)\Big],}{etan} where we have
introduced the function \eqnn{ I(s,\theta)={2\ov
\sqrt{\pi}}{\Gamma(\ft{s}{2}+1)\ov \Gamma(\ft{s}{2}+\ft12 )}
\int\limits_0^{\sinh\theta}  dx\; \big(1\ps x^2\big)^{-1-s/2}.}
 With $\pi I(0,\theta)=2\arctan(\sinh\theta)$ we obtain
\eqnl{ \eta(0,P)=\ft12 \Big\{\eta(0,A)+{2\ov\pi}\,
\zeta(0,A^2)\arctan(\sinh\th)\Big\}\,. }{asyme}

Now, the second term within the curly brackets can be seen to
vanish, since the boundary is a closed manifold of odd
dimensionality. In fact, in our case, $\zeta(0,A^2)=a_1(A^2)=0$,
where $a_1(A^2)$ is a heat kernel coefficient in the notation of
\cite{gilkey} (for details, see Theorem 1.12.2 and Lemma 1.8.2 in
this reference), and we are left with \eqnl{ \eta(0,P)=\ft12
\eta(0,A)\,. }{asym}

As far as $A$ is invertible, this is the main result of this
section, relating the $\eta-$invariant of the Dirac operator to
the same invariant of the boundary operator. Note that the outcome
is the same irrespective of the value of $\theta$, i.e., it holds
both for pure MIT and chiral bag conditions. The first case was
treated before in \cite{woj3}; our result coincides with the one
given in that reference (equation (4.16)), up to an overall factor
$1/2$. This discrepancy seems to be due to an extra factor of $2$
in equations (4.7) and (4.8) in that reference. This extra factor
is inconsistent with equation (4.10), and has seemingly propagated
to Theorem 4.4 in the same paper.

Our result (\ref{asym}) changes sign when the normal to the
boundary points in the opposite direction, since then the
non-diagonal entries in $M$ and $N$ change sign and, as a
consequence, so does the Dirac trace.

As already pointed, (\ref{asym}) gives the whole spectral
asymmetry when the boundary Dirac operator $\gamma_0 A$ is
invertible. In fact, for such cases it was proved in \cite{woj3}
(see also \cite{woj1}) that the asymmetry splits, in the adiabatic
(infinite volume) limit, into the volume contribution plus the
infinite cylinder one. Moreover, reference \cite{müller} shows
that the spectral asymmetry is independent from the size of the
manifold when the boundary value problem is self adjoint, as in
our case. This, together with the vanishing of the volume
contribution in even dimensions, leads to the previous conclusion.

\bigskip

Now, we study the more subtle case of a non-invertible boundary
operator $A$. Then, as can be seen from \refs{t1}, $\an =0$ would
give no extra contribution in the semi-infinite cylinder. However,
in this case, the trace \refs{trace} can differ in a substantial
way from the corresponding one in the collar neighborhood.
As explained in \cite{woj1}, both large $t$ behaviors may
be different, thus giving extra contributions to the asymmetry  in
the collar. This difference in high $t$ behavior is due to the
presence of ``small" eigenvalues, vanishing as the inverse of the
size of the manifold in the adiabatic limit \cite{park}. These
extra contribution can be determined, modulo integers, by using
the arguments in \cite{gilkey,müller,woj2}. To this end, consider
the one-parameter family of differential operators \eqnn{
P_{\al}=P+\frac{2\pi}{\beta}\al\gamma_0,\qquad P_0=P.} These
operators share the same $\al$-independent domain. They are
invertible for $\al \neq 0$ and can be made invertible for all
$\al$ by subtracting the projector on the subspace of small
eigenvalues related to the zero-modes at $\al=0$. This then yields
a new family of operators $P'_{\al}$ and one obtains \eqnn{
\eta(0,P_{\al})=\eta(0,P^{\prime}_{\al})\,\,
\mod\,\mathbb{Z}\mtxt{and} {d\ov d\al}\eta(0,P_\al)={d\ov
d\al}\eta(0,P'_\al).} Then, differentiating with respect to $\al$
one finds \eqngrrl{ {d\ov d\al}\eta(0,P'_\al)&=&
{1\ov \Gamma(\ft{s+1}{2})}{d\ov d\al}
\int\limits_0^\infty\,dt\,t^{\ft{s-1}{2}} \Tr\big(P'_\al
e^{-t{P'_\al}^2}\big)\big\vert_{s=0}} {&=& {1\ov
\Gamma(\ft{s+1}{2})} \int\limits_0^\infty\,dt\,t^{\ft{s-1}{2}}
\Tr\Big[{d P'_\al\ov d\al}\big(1+ 2t{d\ov dt}
\big)e^{-t{P'_\al}^2}\Big]\big\vert_{s=0}} {&=& -{2\pi\, s\ov
\beta\,\Gamma(\ft{s+1}{2})}\int\limits_0^\infty\,dt\,t^{\ft{s-1}{2}}
\Tr\big(\gam_0 e^{-t{P'_\al}^2}\big) +{4\pi\ov
\beta\Gamma(\ft{s+1}{2})} \,\Tr\big(t^{\ft{s+1}{2}}\gam_0
e^{-t{P'_\al}^2}\big)_{t=0}^\infty \Big\vert_{s=0}\,,}{sflux}
where we performed a partial integration to arrive at the last
equation. In addition, we used
$dP'_\al/d\al=\frac{2\pi}{\beta}\gam_0$. Since $P'_{\al}-P_\al$ is
an operator of finite range we may safely skip the prime in the
last line of the above formula. Finally, the very last term in
equation \refs{sflux} can be seen to vanish, which gives, for the
spectral flow (with almost the same calculation as the one
starting with equation \refs{t}, except that no derivative w.r.t
$x_0$ must be taken) \eqnl{ {d\ov
d\al}\eta(0,P^{\prime}_\al)=-{\pi\ov\beta}{\rm Res}|_{s=0}
\Big[\zeta(\ft{s+1}{2},A^2)+ {2\ov\pi}\eta(s\ps 1,A)
\arctan(\sinh\th)\Big]\,.}{sflux2}

Now, the second term can be seen to vanish, since (again with the
notation of \cite{gilkey}), $\sqrt{\pi}\,{\rm Res}|_{s=0}\eta(s\ps
1,A)=2a_{0}(A^2,A)=0$. Moreover, $\sqrt{\pi}\,{\rm
Res}|_{s=0}\zeta(\frac{s\ps
1}{2},A^2)=2a_{0}(A^2)={\beta\ov \pi}$.
Thus, one finally has for the spectral flow, no matter whether $A$
is invertible or not \eqnl{ {d\ov d\al}\eta(0,P_\al)=-1}{sflux3}

So, at variance with the case treated in Theorem 2.3 of reference
\cite{woj2}, the spectral flow doesn't vanish for bag boundary
conditions. As a consequence, the contribution to the asymmetry
coming from boundary zero modes is different from an integer. This
also seems to disagree with the result in \cite{woj3}.
Unfortunately, we were not able to trace the origin of this
discrepancy from the results presented in that reference. However,
we will see, in the next section, an explicit example of how this
works.

\section{The asymmetry in a finite cylinder}\label{section-3}

Here, we consider the simple case of the free Dirac operator on a
finite ``cylinder"  and impose twisted boundary conditions in the
Euclidean time direction ($x^0$ ranges from 0 to $\beta$),
non-local APS boundary conditions at $x^1=0$ and local chiral bag
boundary conditions at $x^1=L$. (Note that twisting the boundary
fiber is equivalent to introducing a constant $A_0$ gauge field in
the Dirac operator).

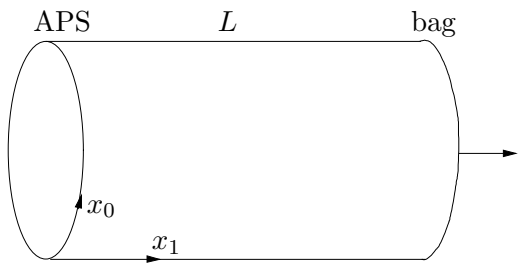
\begin{center}
\setlength{\unitlength}{0.014mm}
\begingroup\makeatletter\ifx\SetFigFont\undefined%
\gdef\SetFigFont#1#2#3#4#5{%
  \reset@font\fontsize{#1}{#2pt}%
  \fontfamily{#3}\fontseries{#4}\fontshape{#5}%
  \selectfont}%
\fi\endgroup%
{\renewcommand{\dashlinestretch}{30}
\begin{picture}(4843,2361)(0,-10)
\put(368,1077){\ellipse{720}{2070}}
\path(413,42)(1448,42)
\blacken\path(1328,12)(1448,42)(1328,72)(1328,12)
\path(368,2112)(3923,2112)
\path(1403,42)(3968,42)
\path(4291,1047)(4831,1047)
\blacken\path(4711,1017)(4831,1047)(4711,1077)(4711,1017)
\path(675,522)(705,650)
\blacken\path(706,526)(705,650)(648,540)(706,526)
\path(3921,2112)(3922,2112)(3923,2113)
    (3926,2114)(3929,2115)(3934,2117)
    (3941,2118)(3948,2120)(3956,2121)
    (3965,2122)(3975,2121)(3985,2119)
    (3996,2116)(4007,2111)(4019,2104)
    (4032,2093)(4046,2080)(4061,2062)
    (4077,2041)(4093,2015)(4108,1989)
    (4122,1962)(4134,1936)(4145,1913)
    (4154,1892)(4162,1874)(4168,1860)
    (4173,1847)(4177,1837)(4180,1827)
    (4183,1818)(4186,1809)(4189,1799)
    (4193,1787)(4197,1773)(4202,1755)
    (4208,1733)(4215,1707)(4223,1675)
    (4232,1639)(4242,1599)(4251,1557)
    (4259,1514)(4266,1474)(4272,1438)
    (4277,1408)(4281,1383)(4283,1363)
    (4285,1347)(4286,1336)(4287,1327)
    (4287,1321)(4287,1315)(4287,1310)
    (4287,1303)(4287,1295)(4287,1283)
    (4287,1268)(4287,1247)(4287,1222)
    (4288,1191)(4288,1154)(4288,1113)
    (4288,1070)(4287,1022)(4285,978)
    (4283,939)(4281,907)(4280,880)
    (4278,860)(4276,845)(4275,833)
    (4274,824)(4273,816)(4271,809)
    (4270,800)(4268,789)(4265,775)
    (4263,756)(4259,732)(4254,703)
    (4249,669)(4243,630)(4236,590)
    (4228,547)(4221,509)(4214,477)
    (4209,453)(4205,435)(4202,422)
    (4199,414)(4198,408)(4197,404)
    (4195,400)(4194,396)(4191,389)
    (4188,379)(4183,365)(4176,346)
    (4167,323)(4157,296)(4146,267)
    (4131,233)(4118,204)(4107,183)
    (4099,168)(4093,159)(4089,153)
    (4086,150)(4083,146)(4079,142)
    (4073,136)(4065,126)(4054,113)
    (4041,97)(4026,80)(4009,64)
    (3995,52)(3983,44)(3972,40)
    (3963,38)(3956,38)(3949,38)
    (3944,39)(3940,41)(3937,42)(3936,42)
\put(248,2210){APS}
\put(2000,2210){$L$}
\put(758,485){$x_0$}
\put(1373,132){$x_1$}
\put(3833,2210){bag}
\end{picture}}
\end{center}

The eigenfunctions of the Dirac operator \refs{op} can be expanded
in eigenfunctions of the boundary operator $A=i\pa_0$, satisfying
twisted boundary conditions in the time-direction with twist
parameter $\al$, $\psi(x_0+\al)=e^{2\pi i\al}\psi(x_0)$, as
follows \eqnl{ \psi=\sum_n \psi_n(x_1)e^{i\an x_0},\qquad \psi_n=
\pmatrix{f_n\cr g_n},}{modeexp1} where the eigenvalues of the
boundary operator read \eqnn{ \an =\ft{2\pi}{\beta}(n+\al),\qquad
n\in\Z.} For definiteness, we will consider $0\leq\al<1$ such that
$a_n\geq 0$ is equivalent to $n\geq 0$ and $a_n<0$ to $n<0$. A
vanishing $\al$ corresponds to periodic boundary conditions, and
$\al=1/2$ to anti-periodic (finite temperature) boundary
conditions. The mode-functions in \refs{modeexp1} fulfill the
simple differential equations \eqnn{ g_n'-a_n g_n=\lam
f_n\mtxt{and} -f_n'-a_n f_n=\lam g_n.} At $x^1=0$, the APS
boundary conditions require \eqnn{ \an\geq 0:\quad
f_n(0)=0\mtxt{and} \an<0:\quad g_n(0)=0\,.} Hence, the
mode-functions have the form \eqnn{ \psi_{n\geq 0}\sim\pmatrix{
\lam \sinh\mu x_1 \cr -\an\sinh\mu x_1-\mu\cosh\mu x_1}\quad,\quad
\psi_{n<0}\sim\pmatrix{ -\an\sinh\mu x_1+\mu\cosh\mu x_1\cr \lam
\sinh\mu x_1}} with $\mu=\sqrt{\an^2-\lam^2}$. On these we must
impose chiral bag boundary conditions at $x_1=L$. The projector
defining these conditions reads \eqnn{ B=\ft12 (1-i\gamf e^{\gamf
\theta } \ns)= \ft12 \pmatrix{1&-e^{\theta } \cr -e^{-\theta
}&1}\,,} and yields the following transcendental equations
\eqngrl{ (\lam e^{-\theta}+\an)\sinh L\mu_n(\lam)+\mu_n(\lam)\cosh
L\mu_n(\lam)=0, && \mtxt{for}n\geq 0} {(\lam e^{\theta} +\an)\sinh
L\mu_n(\lam)-\mu_n(\lam)\cosh L\mu_n(\lam) =0 &&\mtxt{for}n<
0}{modeexp2} for the eigenvalues $\lam(\al)$ of the Dirac operator
on the finite cylinder with APS and bag boundary conditions. With
the evident relation \eqnn{a_{-n-1}(\al)=-a_n(1-\al)} one shows
that the assignment \eqnl{ (n,\al,\theta,\lam)\longrightarrow
(-n-1,1-\al,-\theta,-\lam)}{modeexp3} maps one of the lines of
equation \refs{modeexp2} into the other. Hence, it suffices to
consider the case $n\geq 0$. The contribution of the negative $n$
to spectral functions is taken into account by exploiting the
symmetry \refs{modeexp3}.

Let us first study the asymmetry for $\al\neq 0$, thus excluding
the case of a non-invertible $A$. From the well-known formula
\eqnl{ \sum_\lam \lam^{-s}={1\ov 2\pi i}\oint_\Gamma {dz\ov
z^s}{f'(z)\ov f(z)},}{modeexp4} with $f$ defined by the left hand
side in the first line of equation (\ref{modeexp2}), one obtains

\eqngrl{ \eta(s,P)&=&\frac{1}{2\pi i}
\sum_{n=0}^{\infty}\int\limits_\Gamma {dz\ov z^s} {d\ov
dz}\log\frac{(\an\ps z e^{-\theta})\sinh L\mu_n(z) +\mu_n(z)\cosh
L\mu_n(z)} {(\an\ms z e^{-\theta})\sinh L\mu_n(z) +\mu_n(z)\cosh
L\mu_n(z)}} {&&\quad-\big(\al \rightarrow 1\ms\al,\,\theta
\rightarrow -\theta\big)\,.}{modeexp5} with
$\mu_n(z)=\sqrt{a_n-z^2}$. The curve $\Gamma$ comes from
$\infty+i\epsilon$ to a small semi-circle avoiding the origin and
back to $+\infty-i\epsilon$, surrounding the real positive axis
counterclockwise.
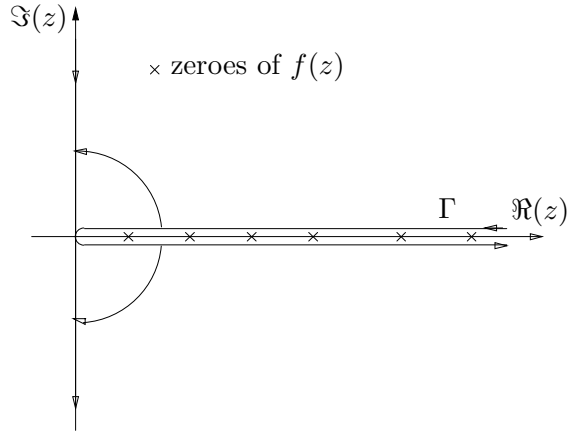
\begin{center}
\setlength{\unitlength}{0.013mm}
\begingroup\makeatletter\ifx\SetFigFont\undefined%
\gdef\SetFigFont#1#2#3#4#5{%
  \reset@font\fontsize{#1}{#2pt}%
  \fontfamily{#3}\fontseries{#4}\fontshape{#5}%

\selectfont}%
\fi\endgroup%
{\renewcommand{\dashlinestretch}{30}
\begin{picture}(5699,4359)(0,-10)
\put(760,1992){\arc{180}{1.5708}{4.7124}}
\put(695,1968){\arc{1720}{0.0860}{1.500}}
\path(780,1155)(652,1155)(773,1095)(771,1155)
\put(695,2004){\arc{1730}{4.6940}{6.1835}}
\path(780,2894)(667,2870)(785,2834)(788,2894)
\blacken\path(705,4212)(675,4332)(645,4212)(705,4212)
\path(675,4332)(675,12) \path(225,1992)(5445,1992)
\path(5325,1962)(5445,1992)(5325,2022)(5325,1962)
\path(5085,2075)(765,2075)
\path(4965,1872)(5085,1902)(4965,1932)(4965,1872)
\path(5085,1910)(765,1910)
\path(4980,2112)(4860,2082)(4980,2052)(4980,2112)
\path(4860,2082)(5040,2082) \path(675,4017)(675,237)
\path(645,357)(675,237)(705,357)(645,357)
\path(645,3687)(675,3567)(705,3687)(645,3687)
\path(675,3567)(675,3702) \path(1170,2037)(1260,1947)
\path(1800,2037)(1890,1947) \path(2430,2037)(2520,1947)
\path(3060,2037)(3150,1947) \path(3960,2037)(4050,1947)
\path(4680,2037)(4770,1947) \path(1170,1947)(1260,2037)
\path(1800,1947)(1890,2037) \path(2430,1947)(2520,2037)
\path(3060,1947)(3150,2037) \path(3960,1947)(4050,2037)
\path(4680,1947)(4770,2037) \path(1440,3747)(1530,3657)
\path(1440,3657)(1530,3747) \put(0,4152){$\Im (z)$}
\put(5130,2172){$\Re(z)$} \put(4387,2187){$\Gamma$}
\put(1650,3657){zeroes of  $f(z)$}
\end{picture}
}
\end{center}

 Now, the contour can be opened to
the imaginary axis, and the circle around the origin can be
shrunken, since the integrand vanishes at $z=0$. After doing so,
one gets \eqnn{ \eta(s,P)=
\frac{1}{i\pi}\sum_{n=0}^\infty\int\limits_0^\infty {dt\ov t^s}
\cos\big(\frac{\pi s}{2}\big) {d\ov dt}\log \frac{
(\an-it\,e^{-\theta})\tanh L\mu_n(it)+\mu_n(it)}
{(\an+it\,e^{-\theta})\tanh L\mu_n(it)+\mu_n(it)}
-\big(\al\rightarrow 1-\al,\,\theta \rightarrow -\theta\big)\,.}

Now, changing variables according to $t=\an u$, one obtains
\eqngr{ \eta(s,P)&=&{1\ov i\pi} \sum_{n=0}^\infty\cos\big({\pi
s\ov 2}\big)\an^{-s} \int\limits_0^\infty{du\ov u^s} {d\ov
du}\log\frac{
(1-iu\,e^{-\theta})\tanh[L\an\sqrt{1+u^2}]+\sqrt{1+u^2}}
{(1+iu\,e^{-\theta})\tanh[L\an\sqrt{1+u^2}]+\sqrt{1+u^2}}}
{&&-(\al\rightarrow 1-\al,\theta\rightarrow -\theta)\,.} The whole
expression can be evaluated at $s=0$, and one obtains the
following simple result for the spectral asymmetry \eqnl{
\eta(0,P)=-\ft12 \left[\zeta_H (0,\al)-\zeta_H (0,1-\al)\right]
=\al-\frac12\,,}{etainv} where $\zeta_{H}$ is the Hurwitz
$\zeta$-function. In particular, the asymmetry vanishes in the
finite temperature case ($\alpha=\frac12$).

As shown in the previous section (equation \refs{asym}), bag
boundary conditions give, in the absence of boundary zero modes, a
contribution $-\ft12 \eta(0,A)$ to the asymmetry. The minus sign
is due to the fact that, at $x_1=L$, the external normal is
$(0,1)$. APS boundary conditions give no contribution at all and,
as a consequence, the total asymmetry is due to bag boundary
conditions. In this case, it can easily be computed in terms of
Hurwitz zeta functions \eqnn{ -\frac12
\eta(0,A)=-\ft12\Big({2\pi\ov\beta}\Big)^{-s} \Big[\sum_{n\geq
0}(n+\al)^{-s}-\sum_{n>0} (n-\al)^{-s})\Big]\Big\vert_{s=0}} which
is seen to reduce to equation \refs{etainv}.

\bigskip

Let us finally study the periodic case, where a boundary zero mode
does exist. The total asymmetry can be obtained as follows: From
the symmetry \refs{modeexp3} it follows, that \eqnn{
(n,\theta,\lam)\longrightarrow (-n,-\theta,-\lam),\qquad n\neq 0}
is a symmetry of the equations \refs{modeexp2}. The contribution
from this modes can be evaluated as in the invertible case, and it
is seen to be $\frac12-\frac{2}{\pi}\arctan{e^{\theta}}$.
Regarding $n=0$, the contribution coming from these modes can be
computed directly in terms of Hurwitz zeta functions, and it gives
$-1+\frac{2}{\pi}\arctan{e^{\theta}}$. So, the sum of both
contributions gives for the total asymmetry $\eta(0,P)=-\frac12$.

This result is again in complete agreement with our general result
in the previous section. In fact, APS boundary conditions do not
contribute to the asymmetry mod $\Z$. The contribution of the
local boundary conditions mod $\Z$ can be gotten from the spectral
flow in equation \refs{sflux}. Hence, \eqnn{
\eta(0,P_0)-\eta(0,P_{1/2}) =\eta(0,P_0) =-{1\ov 2}\,(\mod\, \Z).}

It is interesting to note that in all cases bag boundary
conditions transform the would-be contribution to the index due to
APS boundary conditions into a spectral asymmetry. In fact, the
problem can be easily seen to present no zero modes.

\section{Spectral asymmetry in the disk}
\label{section-4}

In this section, we will study the spectral asymmetry for the free
Dirac operator in a disk, subject to bag boundary conditions at
the radius $R$ and with arbitrary $\th$. Note that we are dealing
with a non-product case. However, we will suggest a plausible
interpretation in terms of our results in \ref{section-2}. The
Dirac operator on the disk, subject to nonlocal APS-conditions has
been carefully analyzed in \cite{diskaps,wirai}. In particular,
the connection to the scattering theory of $P^2$ has been
clarified in \cite{wirai}.

We choose the same chiral representation as in section
\refs{section-1} and take polar coordinates ($r,\varphi$), such
that the free Dirac operator takes the form
\eqnl{
P=i\big(\gamma_r\pa_r+\gamma_\varphi{\pa_\varphi\ov r}\big),
\mtxt{with} \gamma_r=\pmatrix{0&e^{-i\varphi} \cr e^{i\varphi}&
0},\quad \gamma_\varphi=\pmatrix{
  0&-ie^{-i\varphi} \cr ie^{i\varphi}& 0 }.}{dirop}
Here, the angle $\varphi$ is the boundary variable, and $0\leq
r\leq R$ is the outward-growing normal one. With $\ns=\gam_r$ the
projector defining bag boundary conditions at $r=R$ reads \eqnl{
B=\frac12 \big(1-i\gamf e^{\gamf \th}\gamma_r\big) =\frac12
\pmatrix{1&-ie^{\theta-i\varphi}\cr ie^{-\theta+i\varphi}&
1},}{boundop} and the boundary operator at $r=R$ is \eqnn{ A={i\ov
R}\,\pa_\varphi.} We expand the eigenfunctions of the Dirac
operator $P$ in eigenfunctions of the total angular momentum
operator \eqnn{ J={1\ov i}{\pa\ov\pa\varphi}+\frac12 \sigma_3,}
which commutes with both $P$ and $B$, \eqnl{
\psi=\sum_{n=-\infty}^{\infty}c_n \pmatrix{f_{n}(r)e^{in\varphi}
\cr g_{n}(r)e^{i(n+1)\varphi}}.}{eig} The radial mode-functions
are determined by the differential equation $P\psi=\lam\psi$,
together with the bag boundary conditions. The differential
equation implies, \eqnn{ f_n=J_{n}\big(|\lam|r\big) \mtxt{and}
g_n=-i\,\sign(\lam)\, J_{n+1}\big(|\lam|r\big)\,,} where $J_{n}$
is the Bessel function of integer order $n$. The boundary
conditions with boundary operator \refs{boundop} yield \eqnl{
J_{n}\big(|\lam|R\big)-\eth\sign(\lam)
J_{n+1}\big(|\lam|R\big)=0,\qquad n\in \Z \,.}{spec2} Here it is
convenient to consider these conditions for positive and negative
eigenvalues $\lam$ separately. With the help of $J_{-n}(x)=(-)^n
J_n(x)$ they can be written as follows: \eqngr{\lam>0:&&
J_{n}(|\lam|R)-\eth J_{n+1}(|\lam|R)
=J_{n}(|\lam|R)+e^{-\th}J_{n+1}(|\lam|R)=0}
{\lam<0:&&J_{n}(|\lam|R)+\eth J_{n+1}(|\lam|R)
=J_{n}(|\lam|R)-e^{-\th}J_{n+1}(|\lam|R)=0,} where
$n=0,1,2,\dots$. Note these conditions are left invariant by the
replacement \eqnn{ (\lam,\theta)\longrightarrow (-\lam,-\theta).}
Hence, with the help of \refs{modeexp4} the spectral asymmetry is
given by the following contour integral in the complex plane
\eqnn{ \eta(s,P)=\frac{1}{2\pi
i}\sum_{n=0}^{\infty}\int_{\Gamma}dz\,z^{-s}
\frac{d}{dz}\log\left(\frac{J_{n}(zR)-\eth J_{n+1}(zR)}
{J_{n}(zR)+\eth J_{n+1}(zR)}\right)-(\th\rightarrow -\th)\,} where
the contour $\Gamma$ is the same as in \refs{modeexp5}. Again we
deform the path of integration such that we integrate along the
imaginary axis. After doing that, and using the definition of the
modified Bessel functions, \eqnn{ J_n(ix)=i^n I_n(x)\mtxt{and}
J_n(-ix)=(-i)^n I_n(x),} we obtain \eqnl{
\eta(s,P)=\frac{1}{i\pi}\cos\frac{\pi s}{2}
\sum_{n=0}^\infty\;\int\limits_0^\infty dt\,t^{-s}
\frac{d}{dt}\log{I_n(tR)+i\eth I_{n+1}(tR)\ov I_n(tR)-i\eth
I_{n+1}(tR)}-(\th\!\to \!-\th).}{eta2} It is convenient to
separate the contribution from $n=0$, which can be evaluated at
$s=0$ without problems. The corresponding integral gives \eqngrl{
\eta^{n=0}(0,P)&=&\frac{1}{i\pi} \log{I_0(tR)+i\eth I_1(tR)\ov
     I_0(tR)-i\eth I_1(tR)}\Big\vert_0^\infty -(\th\to -\th)}
{&=&\frac{1}{i\pi}\log{1+ie^\th\ov1-ie^\th}-(\th\to -\th)
=\frac{4}{\pi}\Big[{\pi\ov 4}-\arctan e^{-\th}\Big]\,}{zero} For
the remaining subspaces, $n\neq 0$, we add and subtract the first
term in the Debye expansion of the modified Bessel functions.

To this end, we change variables according to $t=nu/R$, so that
\eqnl{ \eta^{n\neq 0}(s,P)=\frac{1}{i\pi} \cos\frac{\pi s}{2}
\sum_{n=1}^{\infty}\left({n\ov R}\right)^{-s}
\!\!\int\limits_0^\infty\! du\,u^{-s}{d\ov du} \log{I_n(nu)+i\eth
I_{n+1}(nu)\ov
     I_n(nu)-i\eth I_{n+1}(nu)}-(\th\!\to\!-\th)\,.}{eta3}
The first term in the Debye expansion of the argument of the
logarithm gives
\eqnn{
\log{I_n(nu)+i\eth I_{n+1}(nu)\ov
     I_n(nu)-i\eth I_{n+1}(nu)}\sim
\log F(u,\theta),\quad F(u,\theta)=-{\sqrt{1+u^2}-1-iu e^{-\th}\ov
\sqrt{1+u^2}-1+iue^{-\th}}\,.} When this is added and subtracted
in equation \refs{eta3}, the subtracted part can be seen to vanish
at $s=0$, since the integrand cancels both at $0$ and $\infty$.
Thus, we are left with \eqnn{ \eta^{n\neq 0}(s,P) ={1\ov i\pi}\cos
\frac{\pi s}{2} \sum_{n=1}^\infty\left({n\ov
R}\right)^{-s}\int\limits_0^\infty du\,u^{-s} {d\ov du}\log
F(u,\theta)-(\th\to-\th)\,,} which yields a finite expression for
$s=0$, \eqnn{ \eta^{n\neq 0}(0,P) ={1\ov i\pi}\zeta_{R}(0)
\log{F(u,\theta)}\Big\vert_0^\infty -(\th\to -\th)\,,} where
$\zeta_{R}$ is the Riemann zeta function. Inserting the values of
$F$ and $\zeta_{R}(0)$ yields \eqnl{ \eta^{n\neq 0}(0,P) ={2\ov
\pi}\Big[\arctan{e^{-\th}}-{\pi\ov 4}\Big]\,.}{nzero} When added
to the contribution in equation \refs{zero}, this gives for the
total asymmetry \eqnl{ \eta(0,P)={2\ov
\pi}\big[\arctan{e^{|\th|}}- \frac{\pi}{4}\Big]\,.}{nzero1} This
is precisely the result predicted by equation \refs{asyme} when
the eigenvalues of $A$ are of the form $\an =n/R$, with $n\in \Z$.
In fact, in this case, $\eta(0,A)=0$ and
$\zeta^{\prime}(0,A^2)=-1$ (This last is evaluated in the subspace
orthogonal to the zero mode). This can be interpreted as follows:
The operator $P$ in equation \refs{dirop} is not of the form
\refs{op}. However, it reduces to such a form (although with an
$r$-dependent $A$) after choosing the eigenfunctions as in
\refs{eig}. Now, due to the different dependence on the tangent
variable of both components in the spinors, $\gamma_0\, A$ never
goes through zero modes and the calculation in the infinite
cylinder seemingly gives the correct value for the asymmetry, even
though this is a non-product example.

\section{Comments}
\label{section-5}

As already pointed out, our result in equation \refs{asym} gives
the answer also in the presence of a constant $A_0$ gauge field,
which can always be eliminated with the only consequence of
twisting the boundary fiber. It can also be shown to hold for a
gauge field such that $A_0=A_0(x_0)$ (independent of the normal
variable) and $A_1=0$ for, then, a gauge transformation can be
performed of the form
$\psi^{\prime}(x_0,x_1)=e^{i\int_0^{x_0}A_0(x)dx}\psi(x_0,x_1)$,
again leading to just a twist in the boundary fiber.

Finally, our result should, in principle, extend to higher even
dimensions. It is clear that this is so in the case of pure MIT
bag boundary conditions ($\th=0$). For chiral bag boundary
conditions, this is not so clear due to to presence of oblique
boundary conditions \cite{pc} (and possible lack of ellipticity)
in the associated second order problem \cite{habklaus, dk, es}.

\section*{Acknowledgements}

We thank H. Falomir and D.V. Vassilevich for carefully reading the
manuscript and for useful suggestions. This work was partially
supported by Fundaci\'on Antorchas and DAAD (grant 13887/1-87),
CONICET (grant 0459/98) and UNLP (grant X298).

\end{document}